\documentclass{article}
\usepackage{spconf,amsmath,graphicx}
\usepackage{microtype} %This fixes the problem of lines extending beyond margin in justified text
\usepackage{mathpazo} %This fixes the problem of lines extending beyond margin in justified text
\usepackage{caption}
\usepackage{subcaption}
\usepackage{url}
\usepackage{enumitem}
\usepackage{longtable}
\usepackage{xtab}
\usepackage{amsmath}

\title{ An overview of techniques for biomarker discovery in voice signal }
\name{Rita Singh, Ankit Shah*, Hira Dhamyal* \thanks{* Second and third author contributed equally}}
\address{Carnegie Mellon University \\ Pittsburgh, PA, USA}

\begin{document}
%\ninept
%
\maketitle
\begin{abstract}

This  paper  reflects on  the  effect  of  several  categories of  medical  conditions  on  human  voice,  focusing  on those that may be hypothesized to have effects on voice, but for which the changes themselves may be subtle enough to have eluded observation in standard analytical examinations of the voice signal. It presents three categories of techniques that can potentially uncover such elusive biomarkers and allow them to be measured and used for predictive and diagnostic purposes. These approaches include proxy techniques, model-based analytical techniques and data-driven AI techniques.

\end{abstract}
\section{Introduction}
\label{sec:intro}

 Based on their effects on human voice, medical conditions that are known to affect humans can be divided into four clear categories. Of these, one category of diseases comprises those that have absolutely no effect on voice, such as certain dermatological conditions, hair-related conditions etc. In contrast to this, the category of conditions that is expected to have the most obvious effects on voice includes diseases that directly affect the structures of the vocal tract -- vocal folds, larynx, glottis, respiratory tract, articulators etc. Examples of such diseases are otolayngological diseases of various etiology. A third category is that of diseases that indirectly affect the processes that drive voice production -- including cognitive, neuromuscular, biomechanical and auditory feedback processes. These conditions cause varied effects on voice, ranging from fairly intense and obvious effects, to very subtle or almost imperceptile ones. This category includes diseases such as those listed in Table 1, and syndromes caused by secondary effects of drugs, intoxicants and other harmful substances. The fourth category -- and the focus of the mechanisms presented in this paper -- comprises diseases for which the  existence of voice changes is hypothesized, but these may not be evident through standard analytical examinations of the voice signal. These include disease subcategories that affect intellectual abilities, visual function, temperament,  personality, behavior etc. For these diseases, biomarkers in voice may be hypothesized to be present, but are elusive and must be searched for -- in effect designed or created -- for use in data-driven applications that attempt to detect the presence of these diseases from voice.

The changes in voice that are alluded to above are in fact biomarker patterns, or biomarkers.
The term biomarker in this context refers to specific patterns of change(s) in the voice signal in the voice signal, that carry information about the health conditions that cause them.
Such changes may be thought of as perturbations or deviations from a hypothetical ``normal'' voice signal, within its frequency, duration, intensity, amplitude and other entities that characterize it. The changes may be wide-raging and coarse in nature, or temporally transient, occurring within micro-durations of the signal. In other words, biomarker patterns may range from being highly perceptible, to completely imperceptible -- to the extent that they may be undetectable even within standard analytical representations of the speech signal.

The rest of this paper introduces three categories of techniques that can potentially uncover such
elusive biomarkers. Such techniques can be thought of as biomarker discovery or feature engineering techniques focused on the design or re-design of biomarker features in voice. The approaches described below include proxy techniques, model-based analytical techniques and data-driven AI techniques.

 % In case of humans, the process of voice production contains the linkage of the different diseases. 

\vspace{2mm}
\hrule
\vspace{2mm}
“© 20XX IEEE. Personal use of this material is permitted. Permission from IEEE must be obtained for all other uses, in any current or future media, including reprinting/republishing this material for advertising or promotional purposes, creating new collective works, for resale or redistribution to servers or lists, or reuse of any copyrighted component of this work in other works.”

\begin{table*}[ht]
    \centering
     \resizebox{2.0\columnwidth}{!}{
    \begin{tabular}{|p{0.27\textwidth} | p{0.73\textwidth} |}
    \hline
\textbf{Health condition}  & \textbf{How it affects voice} \\ \hline
Attention Deficit Hyperactivity Disorder (ADHD) & Prosodic variations in loudness and fundamental frequency \cite{von2021predicting} \\ \hline
 Amyotrophic Lateral Sclerosis (ALS) & Voice tremor, flutter \cite{aronson1992rapid}, incomplete vocal fold closure, dysarthria \cite{chen2005otolaryngologic};  Dystonia, dysarthria  \cite{griffiths1989neurologic}; Low-frequency ($<$ 4 hz) tremor \cite{hlavnivcka2020characterizing} \\ \hline
 Alzheimer's Disease (Dementia) &  Abnormal fundamental frequency, pause and voice-break patterns, reduction in vocal range \cite{meilan2014speech, martinez2021ten}; Dysphonia \cite{woo1992dysphonia} \\ \hline
 Arthritis: Fibromyalgia &  Changes in Jitter, shimmer, harmonic-to-noise ratio, and phonation time \cite{gurbuzler2013voice} \\ \hline
 % Black Lung (Coal Workers' Pneumoconioses) &  Breathless voice \cite{bodenhamer2016breathless}; Phoneme (vowel) duration and intensity perturbations \cite{gilbert1975speech} \\ \hline
%  Bulbar palsy & Dystonia, dysarthria  \cite{griffiths1989neurologic}\\ \hline
% Campylobacter Infection (Campylobacteriosis) & Increased nasality \cite{onodera2002acute}\\ \hline
 Cerebral Palsy & Dysphonia \cite{coombes1991voice}; Breathiness, Asthenia, Roughness,  Strain \cite{miller2013changes} \\ \hline
Cholera  & Husky voice \cite{naidoo1877notes}; high-pitched, asthenia \cite{carpenter1971cholera}; ``Cholera voice'' \cite{humphreys1849cholera} \\ \hline

% (Vibrio cholerae Infection)
% Diphtheria (Corynebacterium diphtheriae Infection) &  Hoarseness \cite{jain2016diphtheria} \\ \hline
% Chronic Traumatic Encephalopathy (CTE) & Aberrant patterns of pauses and disfluencies \cite{krasilshchikova2020speech}; Slurring, dysarthria  \cite{baugh2012chronic} \\ \hline
Congenital Heart Defects & Dystonia, dysarthria  \cite{griffiths1989neurologic}
\\ \hline
% Cerebellar ataxia & Low-frequency ($<$ 4 hz) tremor \cite{hlavnivcka2020characterizing} \\ \hline
Coronavirus Disease 2019 (COVID-19) & Abnormal acoustic measures of pitch, harmonics, jitter, shimmer \cite{asiaee2020voice};  Asymmetric/abnormal vocal fold vibrations \cite{al2021detection, deshmukh2021interpreting} \\ \hline
Diabetes & Roughness, asthenia, breathiness, and strain in high glycemic index subgroups \cite{hamdan2012vocal} \\ \hline
Down Syndrome & Reduced pitch range, higher mean fundamental frequency, reduced jitter \cite{lee2009intonation} \\      \hline
% \cite{ moura2008voice}
%Speech impairment \cite{kent2013speech}; Other spectral changes \cite{albertini2010spectral} 
 % Epstein-Barr Virus (EBV) Infection & Altered resonance patterns, dysphonia \cite{johns2000simultaneous}\\ \hline
 Epilepsy (Temporal Lobe) & Abnormal voice pitch regulation \cite{li2016temporal}\\ \hline
% Essential tremor & Low-frequency ($<$ 4 hz) tremor, Medium tremor (4-7 Hz) \cite{hlavnivcka2020characterizing} \\ \hline
% Musculoskeletal disorders & Voice stress, dysarthria \cite{rubin2007musculoskeletal, roy2008assessment} \\ \hline
% Guillain-Barr{\'e} Syndrome & Asthenia \cite{panosian1993guillain}; Increased nasality \cite{smith1957syndrome, pellegrini2018bilateral}\\ \hline
% Hansen's Disease (Leprosy) & Aphonia \cite{boeckl2021historiography} \\ \hline
Huntington's disease & Low-frequency ($<$ 4 hz) tremor \cite{hlavnivcka2020characterizing}
\\ \hline
Hypertension &  Noisy breathing, voice production difficulty, vocal fatigue \cite{anil2019study} \\ \hline
Hyperthermia (Extreme Heat) & Pressured speech \cite{jamshidi2019hot}; Dysarthria \cite{musselman2013diagnosis} \\ \hline
Hypothermia (Extreme Cold) & Slurred speech \cite{aslam2006hypothermia}\\ \hline
% , hanania1999accidental
Lung Cancer & Hoarseness, GRBAS symptoms, dysphonia \cite{lee2008nature}\\ \hline
% Lymphatic Filariasis & Hoarseness \cite{lim2000patient} \\ \hline
Myasthenia gravis & Dystonia, dysarthria  \cite{griffiths1989neurologic}
\\ \hline
Multiple sclerosis & Low-frequency ($<$ 4 hz) tremor \cite{hlavnivcka2020characterizing}
\\ \hline
% Multiple system atrophy & Low-frequency ($<$ 4 hz) tremor, Medium tremor (4-7 Hz) \cite{hlavnivcka2020characterizing} \\ \hline
Myxoedema & low-pitched, husky, and nasal voice, unusual diction \cite{lloyd1959value}
\\ \hline
% Obesity & Hoarseness, vocal murmur, vocal instability, altered jitter and shimmer, reduced maximum phonation times, voice strangulation at the end of emission\cite{solomon2011obesity, da2011voice} \\ \hline
Parkinson's disease &  Dystonia, dysarthria  \cite{griffiths1989neurologic}; Low-frequency ( $<$ 4 hz) tremor, Medium tremor (4-7 Hz) \cite{hlavnivcka2020characterizing} \\ \hline
%Pseudobulbar palsy & Dystonia, dysarthria  \cite{griffiths1989neurologic} \\ \hline
Pneumonia &  Wet, gurgly voice \cite{marik2003aspiration} \\ \hline
Pulmonary Hypertension & Hoarseness \cite{shah1980hoarseness} \\ \hline
Stress & Breathiness, softness \cite{godin2015physical}, Multiple changes \cite{scherer1986voice}\\ \hline
Tobacco Use & Multiple voice quality changes, high jitter \cite{guimaraes2005health}\\ \hline
% Whitmore's Disease (Melioidosis) &  Hoarseness \cite{inglis2006melioidosis}  \\ \hline
    \end{tabular}
    }
    \caption{Qualitative observations of the effect of some diseases on voice}
    \label{tab:relations}
\end{table*}

\section{Biomarkers and their measurement through proxy techniques}

Proxy techniques may be used for the changes in voice that are human observable, but not easily measurable. For example, many of the correlations established between various diseases and voice changes in the medical literature refer to changes in voice quality. The changes, although subjective, comprise biomarkers that could potentially be used in machine learning systems for prediction of the corresponding diseases from voice. However, the problem is that for the most part, the entities that constitute the set of voice qualities are subjectively specified. For many, methodologies for objective measurement do not exist.

For example, the voice sub-qualities of nasality (or nasalence), roughness, breathiness, asthenia etc. have no objective measures associated with them. They are in fact rated by human experts on standardized clinical rating scales, such as  Voice Rating scale (VRS) \cite{cho2011differences}, Voice Disability Coping Questionnaire (VDCQ) \cite{epstein2009individuals}, 
Consensus Auditory-Perceptual Evaluation of Voice (CAPE-V)  \cite{kempster2009consensus} etc. Examples of such subjective correlations of various diseases to voice changes are listed in Table 1. %For brevity, voice changes are reported in five categories that follow the  minimal protocol prescribed by the European Laryngological Society's minimal for the assessment of voicing aberrations:

%\begin{itemize}[noitemsep,topsep=0pt]
%\item Perception (grade, roughness, breathiness) (Pe)
%\item Videostroboscopy (closure, regularity, mucosal wave and symmetry) (Vi)
%\item Acoustics (jitter, shimmer, F0-range and softest intensity) (Ac)
%\item Aerodynamics (phonation quotient) (Ae)
%\item Subjective rating by the patient.  (Su)
%\end{itemize}

%This protocol is elaborated in \cite{dejonckere2001basic} and we use this in the table below. Broader medical assessment protocols separate out Perception (Pe) and Acoustics (Ac), and compress Vi, Ae etc. in the single category of Phonation (Ph), and . In addition to these they include other broad categories of changes such as Articulation (Ar), Evolution (of speech, such as stuttering, pauses etc.) (Ev) etc. When prominent, these are reported in the table below as well. 

\subsection{Proxy features}

Features that are subjectively rated and do not have specific methods to objectively measure them can be measured through proxy. The term ``proxy'' here refers to the act of using (or creating) replacement features that can be measured instead of the original ones. For this to be viable, the proxy features must be highly correlated to the subjective features that we desire to measure. There are two potential mechanisms to create such proxy features: the use of physical models of voice production to produce signals that can be more easily measured, and the use of AI mechanisms for transfer learning that can generate measurable features that exhibit the same patterns as the subjective features they proxy for.

\subsubsection{Proxy features from models of voice production: measurement through emulation}

As an example, we consider physical models that can generate just one specific aspect of a continuous speech signal -- the set of phonated sounds embedded in it. The idea is to use such models to generate or approximate the actual motion of the vocal folds during the process of phonation, producing a glottal flow signal that has characteristics (or specific voice sub-qualities) that are similar to a give recorded speech signal. Once this is achieved, the parameters of the model used can proxy for the speech quality characteristics of the original signal, that may not have been directly measurable. We can also call this process ``measurement through emulation.'' Examples of physical models of phonation that can be used include the 1-mass, mass-spring model \cite{lucero2013modeling},  the 2-mass, mass-spring model \cite{lucero1993dynamics} etc.

For a given recording, the parameters of these models are derivable through the ADLES algorithm, that minimizes the squared error between a glottal flow signal estimated through inverse filtering and the glottal flow signal generated through the model. The discriminability of such features is evident from the highly characteristic patterns exhibited by the corresponding model in its phase space. For example, while we know that there are changes in voice in response to various diseases of the vocal folds, and may have observed changes in voice as a result of covid, it is hard to characterize, identify or measure the exact changes in spectrographic and other signal representations. However, we can also see that the vocal fold movements (phonation process) would be affected by these diseases, and be highly correlated to the actual vocal fold oscillations as an affected person speaks. This is borne out by the phase space patterns exhibited by a 1-mass model, as shown in Fig. 1. We can therefore use the corresponding parameter values (and even other measurements that pertain to the phase space trajectories shown) to build predictors for the underlying conditions. The model parameters are then the proxy features we are looking for.

\begin{figure}[t]
\centering
\begin{subfigure}[t]{0.154\textwidth}
\centering
\includegraphics[width=\textwidth]{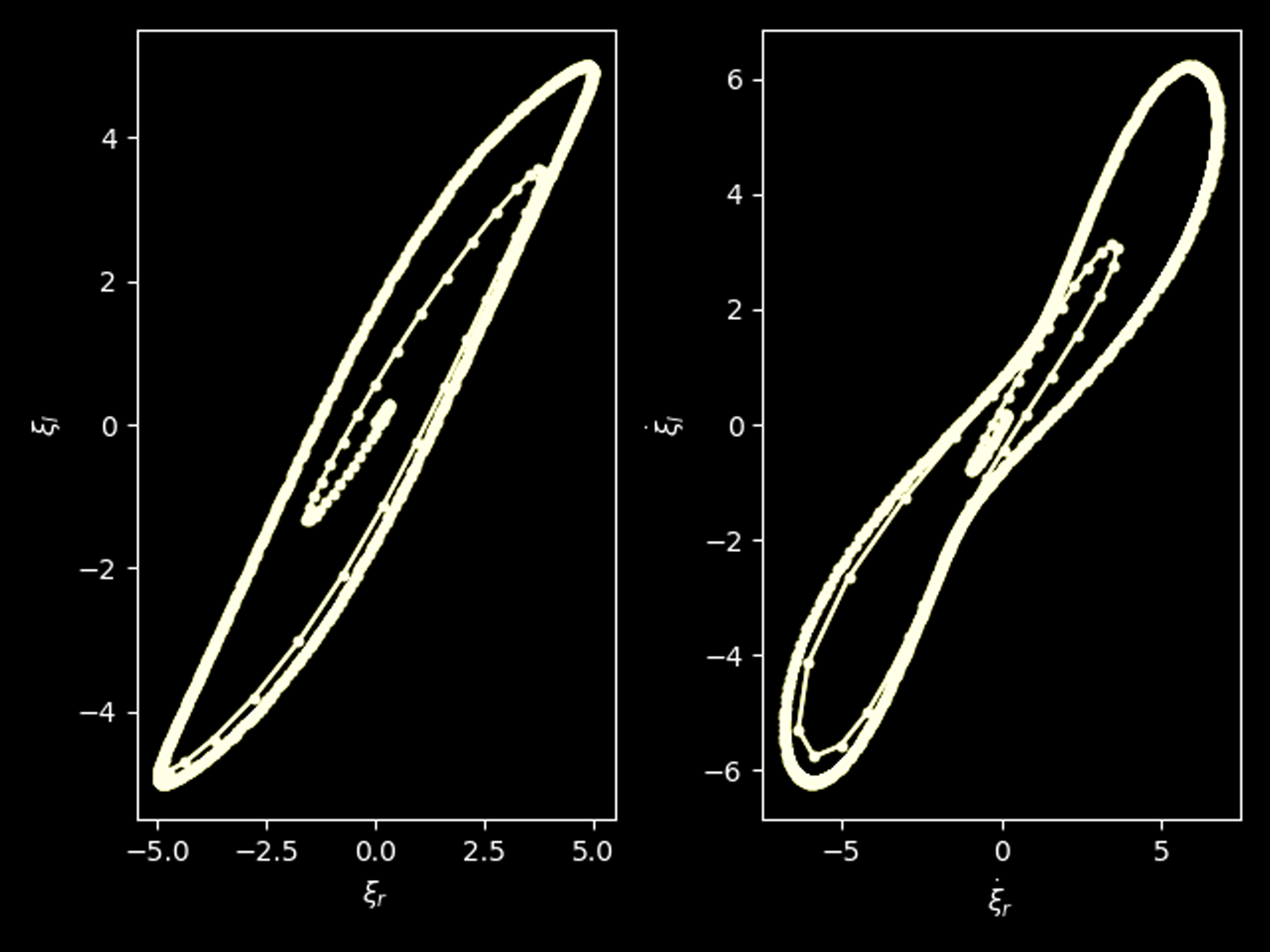}
\caption{Normal voice
}\label{fig:p1}
\end{subfigure}
\begin{subfigure}[t]{0.158\textwidth}
\centering
\includegraphics[width=\textwidth]{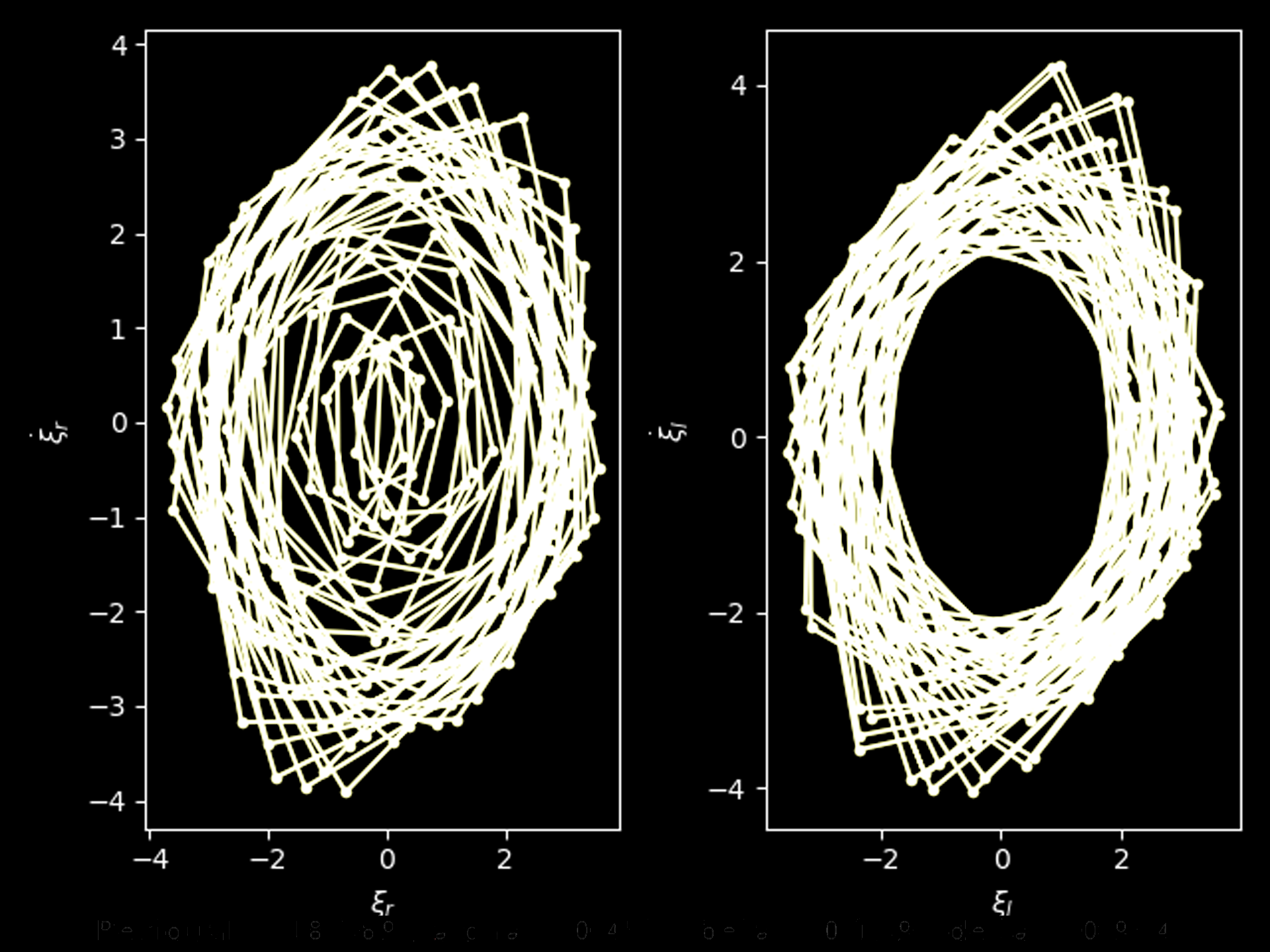}
\caption{Covid-19 ($>$ 7 days after +ve test)
}\label{fig:p5}
\end{subfigure}
\begin{subfigure}[t]{0.158\textwidth}
\centering
\includegraphics[width=\textwidth]{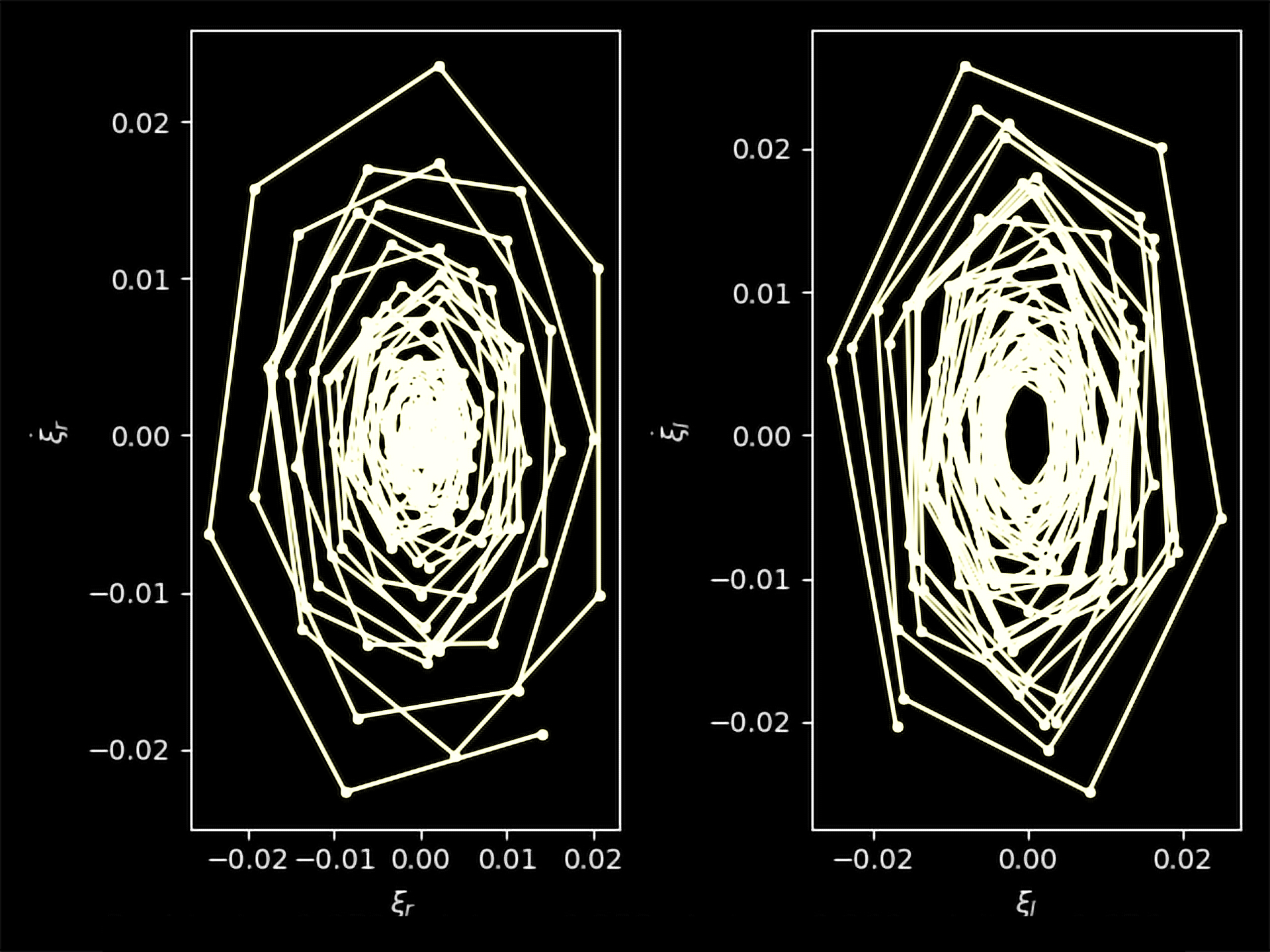}
\caption{Covid-19 ($<$ 7 days after +ve test)
}\label{fig:p6}
\end{subfigure}
\caption{Phase space trajectories of the 1-mass model (for the extended vowel /aa/) estimated to emulate the bilateral vocal fold vibrations of people with normal voice and Covid-19.}
%\title{Phase space trajectories of a 1-mass model of phonation} 
\label{fig:phase}
\end{figure}

\subsubsection{Proxy features from neural networks: measurement through correlation}

Proxy features can also be derived using neural networks and other classifiers that learn to perform classification tasks on the signals within which we desire to identify or measure biomarkers. When trained to perform auxiliary tasks that are equivalent to those that must be actually performed using objective measures of the biomarker, the scores generated through the auxiliary classifiers act as proxy features for the biomarkers in question. For example, it is easy to identify, but difficult to measure changes in the ``nasality'' of speech signals. However, it is relatively easy to identify and isolate nasal and non-nasal phonemes in speech, and build classifiers to discriminate between them. When properly trained, an accurate classifier would generate scores that are discriminative of nasality characteristics. If this were not so, the classifier would not be able to accurately perform the task of discriminating between nasal and non-nasal phonemes. Once trained, the classifier can be used to generate scores for training newer predictors of underlying conditions based on nasality.  that we wish to measure.

\section{AI systems for biomarker discovery}
With the advancements in the voice profiling techniques, it has become evident that for a large set of diseases for which correlations to voice were not known to exist, the presence of such correlations can in fact be hypothesized and scientifically supported. Since for these, we clearly know that biomarkers are not perceptible or evident within standard analytical representations of the voice signal, we must devise techniques to discover them or create them in appropriate mathematical spaces within which they can be shaped and measured.

\begin{figure}[th]
\includegraphics[width=\hsize]{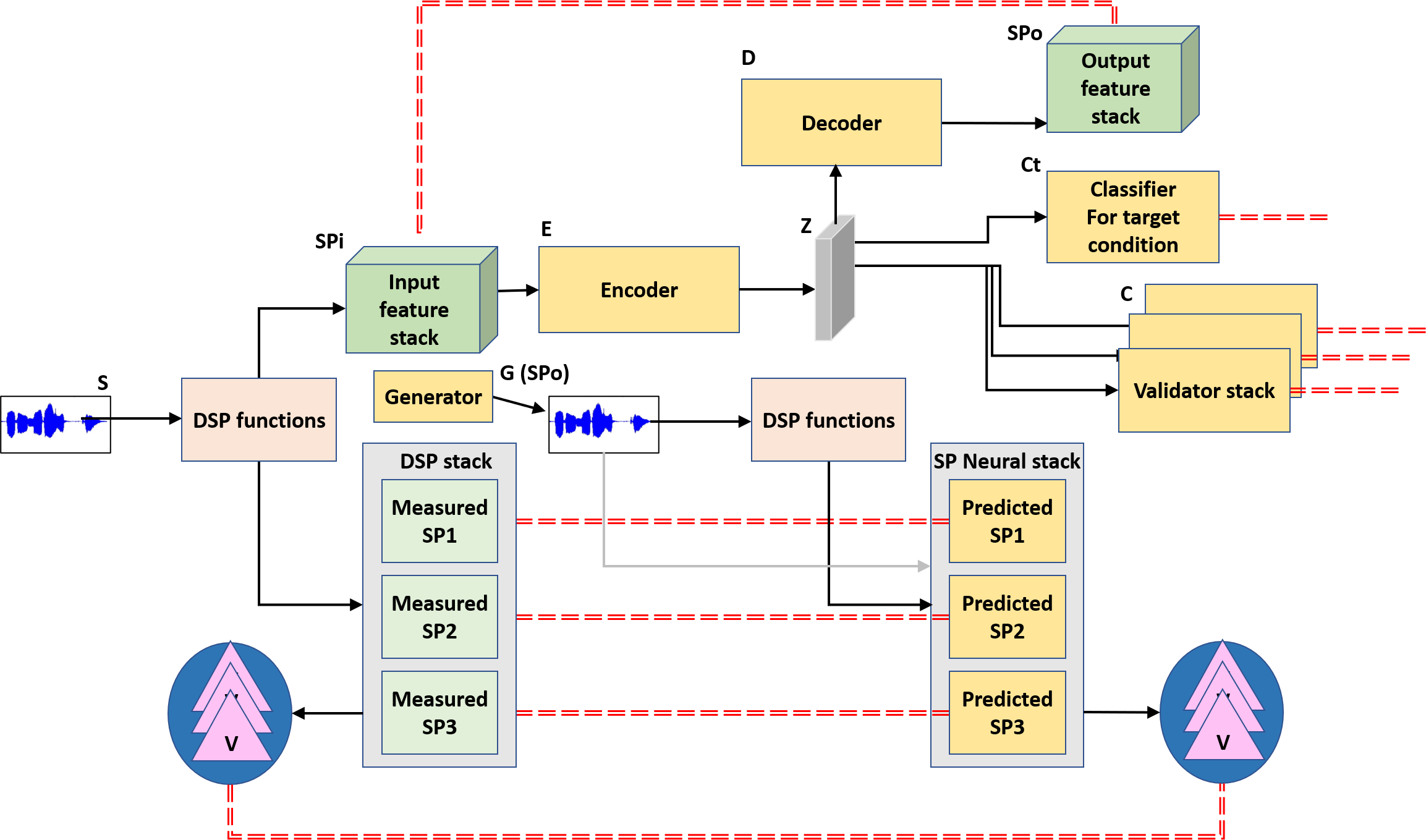}
\caption{Schematic diagram of ABCDE neural model for feature discovery. The yellow boxes represent appropriately architectured neural networks.}
\title{A system for feature discovery}
\label{fig:profilingsystem}
\end{figure}

One example of such a biomarker discovery system/framework is illustrated in Fig. \ref{fig:profilingsystem}. This represents a generic setup which we call the ABCDE framework (\textbf{A}utoencoder based \textbf{B}iomarker \textbf{C}reation and \textbf{D}iscovery \textbf{E}ngine). Its exact formulation can vary from application to application. In this framework, a speech signal is first converted into a numerical representation that is hypothesized to contain the biomarker related information that we seek to uncover, or extract. The representation is then  ``projected'' into a neural kernel space, where we can impose objective criteria on it that are tailored to the specific biomarker.

Viewing Fig.\ref{fig:profilingsystem} from the left, from a given voice signal \textbf{S}, several types of signal representations are obtained using digital signal processing algorithms (DSP). Thereafter, one of those (\textbf{SPi}) is chosen as the substrate for the target feature to be created (carrying the relevant biomarker). The other representations (\textbf{SP1}, \textbf{SP2},...) serve as ``control'' representations, and are placed within the DSP stack shown in the figure \ref{fig:profilingsystem}.

The cuboid for \textbf{SPi} represents a \textit{stack} of the same ``type'' of DSP features, e.g. a stack of spectrograms obtained at different time-frequency resolutions, or a correlogram. These are input into a neural network \textbf{E}  (e.g. a convolutional neural network), which can be viewed as an ``encoder'' that transforms them into a kernel space, yielding a latent feature representation \textbf{Z}. Within this space, different constraints (including those based on prior knowledge) can be imposed on \textbf{Z}, so that it has the desired properties of the biomarker for the targeted health condition. 
One obvious property it must possess is that it must be discriminative for the underlying medical condition it is expected to encode. 

To ensure that the information present in the input representation preserved  in the process of transformation into the kernel space, a decoder \textbf{D} is trained to reconstruct the original representation from \textbf{Z}, while minimizing the loss between the reconstructed and original representation. All loss functions are represented by red double-lines in Fig. \ref{fig:profilingsystem}. In the training process, which is that of parameter optimization  of the aggregate neural framework through gradient descent, such a loss function would be minimized.

To ensure that \textit{only} the information in the original input representation is engineered for the feature at hand, the same latent representation \textbf{Z} is input into a generator \textbf{G} that can recreate the original speech signal from it (as \textbf{G}(\textbf{Z})). Alternatively, the \textbf{G} can act on the output of the decoder, \textbf{SPo} (as \textbf{G}(\textbf{SPo})) to generate the voice signal. Time-domain signals with same information content can be different in voice acoustics and thus, more robust losses are defined based on comparisons of DSP features from the original signal, to those predicted from the voice signal generated by \textbf{G}. Alternatively, the DSP features can be ``simulated'' by a neural stack, as shown in the figure. An additional level of detail that allows voice quality features to play a role in this process of discovery is introduced by specifying rough functional relationships between DSP features and various voice qualities \textbf{V}. Losses based on direct comparisons between these can also play a part in the learning process.

The entire framework is simultaneously optimized, but it is conceivable to express this process of discovery within more sophisticated AI learning frameworks that prioritize interpretability and can be trained in parts. Here, \textbf{Z} represents the ``discovered''  feature carrying the biomarker whose existence was hypothesized. 

\section{Conclusions}

The biomarker discovery mechanisms given above are generic ones, and may have varied formulations in different settings.
In contrast to traditional features derived from audio signals for use with machine learning algorithms, these features are likely to perform better with lesser training data, since they are designed to be relatively more discriminating and less ambiguous for the specific disease for which they are designed. Such features are useful in many ways. For example, they can be used in applications that serve as diagnostic aids in clinical settings, to build tools for early-detection of certain diseases, for self-monitoring of health by disabled, elderly and under-resourced people, etc.

\section{Acknowledgement}
This material is based upon work supported by the Defence Science and Technology Agency,  Singapore under contract number A025959. Its content does not reflect the position or policy of DSTA and no official endorsement should be inferred. 

% \subsection{Existence of biomarkers in speech}
% \input{existence_biomarkers}

% \section{Subjective features and their objective proxies}
% \input{subjective_objective}

% \section{Pushing the envelope: Hypothesizing biomarkers that do not seem
% to exist}
% \input{pushing_envelope}
% \input{feature_engineering.tex}

% \input{application.tex}

\ninept
\bibliographystyle{IEEE}
\bibliography{refs}

\begin{thebibliography}{10}

\bibitem{von2021predicting}
Georg~G von Polier, Eike Ahlers, Julia Amunts, Joerg Langner, Kaustubh~R Patil,
  Simon~B Eickhoff, Florian Helmhold, and Daina Langner,
\newblock ``Predicting adult attention deficit hyperactivity disorder (adhd)
  using vocal acoustic features,''
\newblock {\em medRxiv}, 2021.

\bibitem{aronson1992rapid}
Arnold~E Aronson, William~S Winholtz, Lorraine~Olson Ramig, and Sandra~R
  Silber,
\newblock ``Rapid voice tremor, or “flutter,” in amyotrophic lateral
  sclerosis,''
\newblock {\em Annals of Otology, Rhinology \& Laryngology}, vol. 101, no. 6,
  pp. 511--518, 1992.

\bibitem{chen2005otolaryngologic}
Anton Chen and C~Gaelyn Garrett,
\newblock ``Otolaryngologic presentations of amyotrophic lateral sclerosis,''
\newblock {\em Otolaryngology—Head and Neck Surgery}, vol. 132, no. 3, pp.
  500--504, 2005.

\bibitem{griffiths1989neurologic}
Chester Griffiths and I~David Bough~Jr,
\newblock ``Neurologic diseases and their effect on voice,''
\newblock {\em Journal of Voice}, vol. 3, no. 2, pp. 148--156, 1989.

\bibitem{hlavnivcka2020characterizing}
Jan Hlavnička, Tereza Tykalová, Olga Ulmanová, Petr Dušek, Dana Horáková,
  Evžen Růžička, Jiří Klempíř, and Jan Rusz,
\newblock ``Characterizing vocal tremor in progressive neurological diseases
  via automated acoustic analyses,''
\newblock {\em Clinical Neurophysiology}, vol. 131, no. 5, pp. 1155--1165,
  2020.

\bibitem{meilan2014speech}
Juan Jos{\'e}~G Meil{\'a}n, Francisco Mart{\'\i}nez-S{\'a}nchez, Juan Carro,
  Dolores~E L{\'o}pez, Lymarie Millian-Morell, and Jos{\'e}~M Arana,
\newblock ``Speech in alzheimer's disease: can temporal and acoustic parameters
  discriminate dementia?,''
\newblock {\em Dementia and Geriatric Cognitive Disorders}, vol. 37, no. 5-6,
  pp. 327--334, 2014.

\bibitem{martinez2021ten}
Israel Mart{\'\i}nez-Nicol{\'a}s, Thide~E Llorente, Francisco
  Mart{\'\i}nez-S{\'a}nchez, and Juan Jos{\'e}~G Meil{\'a}n,
\newblock ``Ten years of research on automatic voice and speech analysis of
  people with alzheimer's disease and mild cognitive impairment: A systematic
  review article,''
\newblock {\em Frontiers in Psychology}, vol. 12, pp. 645, 2021.

\bibitem{woo1992dysphonia}
Peak Woo, Janina Casper, Raymond Colton, and David Brewer,
\newblock ``Dysphonia in the aging: physiology versus disease,''
\newblock {\em The Laryngoscope}, vol. 102, no. 2, pp. 139--144, 1992.

\bibitem{gurbuzler2013voice}
Levent Gurbuzler, Ahmet Inanir, Kursat Yelken, Sema Koc, Ahmet Eyibilen, and
  Ismail~Onder Uysal,
\newblock ``Voice disorder in patients with fibromyalgia,''
\newblock {\em Auris Nasus Larynx}, vol. 40, no. 6, pp. 554--557, 2013.

\bibitem{coombes1991voice}
Kay Coombes,
\newblock ``Voice in people with cerebral palsy,''
\newblock in {\em Voice Disorders and their Management}, pp. 202--237.
  Springer, 1991.

\bibitem{miller2013changes}
Nick Miller, Lindsay Pennington, Sheila Robson, Ella Roelant, Nick Steen, and
  Eftychia Lombardo,
\newblock ``Changes in voice quality after speech-language therapy intervention
  in older children with cerebral palsy,''
\newblock {\em Folia Phoniatrica et Logopaedica}, vol. 65, no. 4, pp. 200--207,
  2013.

\bibitem{naidoo1877notes}
T~Narrain~Sawmy Naidoo,
\newblock ``Notes of cases of cholera treated by sulphurous acid,''
\newblock {\em The Indian medical gazette}, vol. 12, no. 8, pp. 219, 1877.

\bibitem{carpenter1971cholera}
CC~Carpenter,
\newblock ``Cholera: diagnosis and treatment.,''
\newblock {\em Bulletin of the New York Academy of Medicine}, vol. 47, no. 10,
  pp. 1192, 1971.

\bibitem{humphreys1849cholera}
Frederick Humphreys,
\newblock {\em The Cholera and Its Homoeopathic Treatment},
\newblock Radde, 1849.

\bibitem{asiaee2020voice}
Maral Asiaee, Amir Vahedian-Azimi, Seyed~Shahab Atashi, Abdalsamad Keramatfar,
  and Mandana Nourbakhsh,
\newblock ``Voice quality evaluation in patients with covid-19: An acoustic
  analysis,''
\newblock {\em Journal of Voice}, 2020.

\bibitem{al2021detection}
Mahmoud Al~Ismail, Soham Deshmukh, and Rita Singh,
\newblock ``Detection of covid-19 through the analysis of vocal fold
  oscillations,''
\newblock in {\em ICASSP 2021-2021 IEEE International Conference on Acoustics,
  Speech and Signal Processing (ICASSP)}. IEEE, 2021, pp. 1035--1039.

\bibitem{deshmukh2021interpreting}
Soham Deshmukh, Mahmoud Al~Ismail, and Rita Singh,
\newblock ``Interpreting glottal flow dynamics for detecting covid-19 from
  voice,''
\newblock in {\em ICASSP 2021-2021 IEEE International Conference on Acoustics,
  Speech and Signal Processing (ICASSP)}. IEEE, 2021, pp. 1055--1059.

\bibitem{hamdan2012vocal}
Abdul-latif Hamdan, Jad Jabbour, Jihad Nassar, Iyad Dahouk, and Sami~T Azar,
\newblock ``Vocal characteristics in patients with type 2 diabetes mellitus,''
\newblock {\em European Archives of Oto-Rhino-Laryngology}, vol. 269, no. 5,
  pp. 1489--1495, 2012.

\bibitem{lee2009intonation}
Mary~T Lee, Jude Thorpe, and Jo~Verhoeven,
\newblock ``Intonation and phonation in young adults with down syndrome,''
\newblock {\em Journal of Voice}, vol. 23, no. 1, pp. 82--87, 2009.

\bibitem{li2016temporal}
Weifeng Li, Ziyi Chen, Nan Yan, Jeffery~A Jones, Zhiqiang Guo, Xiyan Huang,
  Shaozhen Chen, Peng Liu, and Hanjun Liu,
\newblock ``Temporal lobe epilepsy alters auditory-motor integration for voice
  control,''
\newblock {\em Scientific reports}, vol. 6, no. 1, pp. 1--13, 2016.

\bibitem{anil2019study}
HT~Anil, N~Lasya Raj, and Nikitha Pillai,
\newblock ``A study on etiopathogenesis of vocal cord paresis and palsy in a
  tertiary centre,''
\newblock {\em Indian Journal of Otolaryngology and Head \& Neck Surgery}, vol.
  71, no. 3, pp. 383--389, 2019.

\bibitem{jamshidi2019hot}
Nazila Jamshidi and Andrew Dawson,
\newblock ``The hot patient: acute drug-induced hyperthermia,''
\newblock {\em Australian prescriber}, vol. 42, no. 1, pp. 24, 2019.

\bibitem{musselman2013diagnosis}
Megan~E Musselman and Suprat Saely,
\newblock ``Diagnosis and treatment of drug-induced hyperthermia,''
\newblock {\em American Journal of Health-System Pharmacy}, vol. 70, no. 1, pp.
  34--42, 2013.

\bibitem{aslam2006hypothermia}
Ahmed~Faraz Aslam, Ahmad~Kamal Aslam, Balendu~C Vasavada, and Ijaz~A Khan,
\newblock ``Hypothermia: evaluation, electrocardiographic manifestations, and
  management,''
\newblock {\em The American journal of medicine}, vol. 119, no. 4, pp.
  297--301, 2006.

\bibitem{lee2008nature}
Clare~F Lee, Paul~N Carding, and Mike Fletcher,
\newblock ``The nature and severity of voice disorders in lung cancer
  patients,''
\newblock {\em Logopedics Phoniatrics Vocology}, vol. 33, no. 2, pp. 93--103,
  2008.

\bibitem{lloyd1959value}
WH~Lloyd,
\newblock ``Value of the voice in diagnosis of myxoedema in the elderly,''
\newblock {\em British medical journal}, vol. 1, no. 5131, pp. 1208, 1959.

\bibitem{marik2003aspiration}
Paul~E Marik and Danielle Kaplan,
\newblock ``Aspiration pneumonia and dysphagia in the elderly,''
\newblock {\em Chest}, vol. 124, no. 1, pp. 328--336, 2003.

\bibitem{shah1980hoarseness}
KD~Shah, KH~Ayyer, and UK~Shah,
\newblock ``Hoarseness of voice—a presenting manifestation of primary
  pulmonary hypertension,''
\newblock {\em Indian Journal of Otolaryngology}, vol. 32, no. 2, pp. 35--36,
  1980.

\bibitem{godin2015physical}
Keith~W Godin and John~HL Hansen,
\newblock ``Physical task stress and speaker variability in voice quality,''
\newblock {\em EURASIP Journal on Audio, Speech, and Music Processing}, vol.
  2015, no. 1, pp. 1--13, 2015.

\bibitem{scherer1986voice}
Klaus~R Scherer,
\newblock ``Voice, stress, and emotion,''
\newblock in {\em Dynamics of stress}, pp. 157--179. Springer, 1986.

\bibitem{guimaraes2005health}
Isabel Guimar{\~a}es and Evelyn Abberton,
\newblock ``Health and voice quality in smokers: an exploratory
  investigation,''
\newblock {\em Logopedics Phoniatrics Vocology}, vol. 30, no. 3-4, pp.
  185--191, 2005.

\bibitem{cho2011differences}
Shin-Woong Cho, Chang~Shik Yin, Young-Bae Park, and Young-Jae Park,
\newblock ``Differences in self-rated, perceived, and acoustic voice qualities
  between high-and low-fatigue groups,''
\newblock {\em Journal of Voice}, vol. 25, no. 5, pp. 544--552, 2011.

\bibitem{epstein2009individuals}
Ruth Epstein, Shashivadan~P Hirani, Jan Stygall, and Stanton~P Newman,
\newblock ``How do individuals cope with voice disorders? introducing the voice
  disability coping questionnaire,''
\newblock {\em Journal of Voice}, vol. 23, no. 2, pp. 209--217, 2009.

\bibitem{kempster2009consensus}
Gail~B Kempster, Bruce~R Gerratt, Katherine~Verdolini Abbott, Julie
  Barkmeier-Kraemer, and Robert~E Hillman,
\newblock ``Consensus auditory-perceptual evaluation of voice: development of a
  standardized clinical protocol,''
\newblock {\em American Journal of Speech-Language Pathology (AJSLP)}, vol. 18,
  no. 2, pp. 124--132, 2009.

\bibitem{lucero2013modeling}
Jorge~C Lucero and Jean Schoentgen,
\newblock ``Modeling vocal fold asymmetries with coupled van der pol
  oscillators,''
\newblock in {\em Proceedings of Meetings on Acoustics ICA2013}. Acoustical
  Society of America, 2013, vol.~19, p. 060165.

\bibitem{lucero1993dynamics}
Jorge~C Lucero,
\newblock ``Dynamics of the two-mass model of the vocal folds: Equilibria,
  bifurcations, and oscillation region,''
\newblock {\em The Journal of the Acoustical Society of America}, vol. 94, no.
  6, pp. 3104--3111, 1993.

\end{thebibliography}

\end{document}